\documentclass{article}

\usepackage{PRIMEarxiv}
\usepackage[utf8]{inputenc} 
\usepackage[T1]{fontenc}    
\usepackage{hyperref}       
\usepackage{subcaption}
\usepackage{url}            
\usepackage{booktabs}       
\usepackage{amsfonts, amsmath}       
\usepackage{nicefrac}       
\usepackage{microtype}      
\usepackage{lipsum}
\usepackage{fancyhdr}       
\usepackage{graphicx}       
\usepackage{cleveref}
\usepackage{float}
\usepackage{multirow}
\usepackage{xcolor}
\usepackage{fdsymbol}
\usepackage{float}

\restylefloat{table}
\graphicspath{{media/}}     

\title{Inter-Beat Interval Estimation with Tiramisu Model:
A Novel Approach with Reduced Error
\thanks{Code available at: \textcolor{blue}{\url{https://github.com/Arefeen06088/IBI_Tiramisu}}} 
}

\author{
  Asiful Arefeen \\
  EECS \\
  Washington State University \\
  \texttt{asiful.arefeen@wsu.edu} \\
   \And
  Ali Akbari \\
  BME \\
  Texas A\&M University \\
  \texttt{aliakbari@tamu.edu} \\
     \And
  Seyed Iman Mirzadeh \\
  EECS \\
  Washington State University \\
  \texttt{seyediman.mirzadeh@wsu.edu} \\
     \And
  Roozbeh Jafari \\
  BME, CSE and ECE \\
  Texas A\&M University \\
  \texttt{rjafari@tamu.edu} \\
     \And
  Behrooz A. Shirazi \\
  EECS \\
  Washington State University \\
  \texttt{shirazi@wsu.edu} \\
     \And
  Hassan Ghasemzadeh \\
  EECS \\
  Washington State University \\
  \texttt{hassan.ghasemzadeh@wsu.edu} \\
}

\begin{document}
\maketitle

\begin{abstract}
Inter-beat interval (IBI) measurement enables estimation of heart-tare variability (HRV) which, in turns, can provide early indication of potential cardiovascular diseases. However, extracting IBIs from noisy signals is challenging since the morphology of the signal is distorted in the presence of the noise. Electrocardiogram (ECG) of a person in heavy motion is highly corrupted with noise, known as motion-artifact, and IBI extracted from it is inaccurate. As a part of remote health monitoring and wearable system development, denoising ECG signals and estimating IBIs correctly from them have become an emerging topic among signal-processing researchers. Apart from conventional methods, deep-learning techniques have been successfully used in signal denoising recently, and diagnosis process has become easier, leading to accuracy levels that were previously unachievable. We propose a deep-learning approach leveraging tiramisu autoencoder model to suppress motion-artifact noise and make the R-peaks of the ECG signal prominent even in the presence of high-intensity motion. After denoising, IBIs are estimated more accurately expediting diagnosis tasks. Results illustrate that our method enables IBI estimation from noisy ECG signals with SNR up to -30dB with average root mean square error (RMSE) of 13 milliseconds for estimated IBIs. At this noise level, our error percentage remains below 8\% and outperforms other state of the art techniques. 
\end{abstract}


\section{Introduction}
Cardiovascular diseases (CVDs) have been a major concern for human health for a long time and touted as the leading cause of death \cite{CDC_01}. World Health Organization (WHO) declared that CVDs were responsible for the death of almost 17.9 million people in 2016 - 31\% of all global deaths. 85\% of these deaths are due to heart attack and stroke \cite{Aa01}. CVDs can also cause permanent or temporary disabilities and reduce life-quality \cite{Ab02}. As a result, numerous research in this field with a view to preventing people from cardiac diseases have been carried out for the past decades. For in-time CVD diagnosis, it is vital to contentiously monitor HRV parameters that requires accurate estimation of IBI \cite{Harju2018MonitoringOH, Ahmed2010HeartRA}.

\begin{figure}[!htbp]
  \centering
  \includegraphics[scale = 0.425]{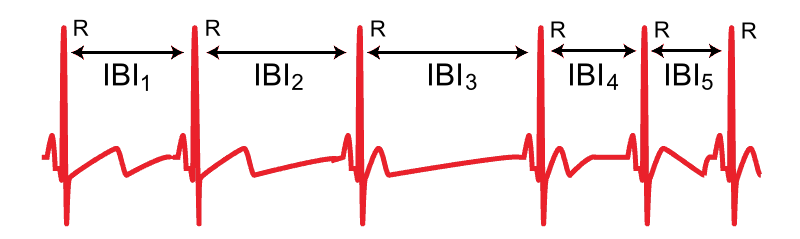}
  \caption{R-R interval or IBI in an ECG.} 
  \label{fig:ibi}
\end{figure}

IBI is one of the most important parameters that can be extracted from ECG and photoplethysmography (PPG) signals. IBI in an ECG signal is the time interval between two consecutive beats as shown in Figure~\ref{fig:ibi}. It is generally measured in units of milliseconds and in normal heart function, each IBI value varies from beat to beat. This natural variation is known as HRV. However, certain cardiac conditions may cause the individual IBI values to become nearly constant, resulting in lower HRV values. This can happen, for example, during periods of exercise as the heart rate (HR) increases and the beats become more regular. Certain illnesses can cause the HR to increase and become uniform as well, such as when a subject is afflicted by an infection. In fact, IBI and HRV can be used as indicators of degraded cardiac system and can be early indicator of certain cardiac diseases \cite{Chessa2002RoleOH}. 

Wearable sensors provide opportunity for continuous and convenient measurement of health parameters such as IBI. For the last decade, there has been immense development in the domain of wearable devices for physiological parameter monitoring, disease diagnosis and early prevention. Sleep apnea monitoring \cite{Ac03,Ad04}, cardiac anomalies or arrhythmia detection/classification \cite{Ae05,Af06,Qayyum2019ECGHC}, HRV estimation \cite{Ag07}, ECG monitoring \cite{Ah08}, and ophthalmic disease diagnosis \cite{Islam2019SourceAC} are some of the tasks worth mentioning. 

Although PPG sensors embedded in various wearables such as smart watches and smart rings enable convenient and continuous measurement of IBI, extracting IBI values from PPG is challenging as PPG is highly prone to noise. On the other hand, ECG signal that is encrypted with Gaussian out-of-band noise can be denoised easily using different types of finite impulse response (FIR) filters and empirical mode decomposition (EMD) techniques \cite{Chang2011GaussianNF}. However, an ECG signal that is encrypted with motion artifacts cannot be dealt easily as the frequency spectrum of some of these noises (1-10 Hz) \cite{Kher15} overlaps with that of the PQRST contents (0-50 Hz) \cite{Tereshchenko2015FrequencyCA,Kirst2011UsingDF}. Such situations emerge specially when the subject is in constant and/or periodic motion or performs exercise, where extracting IBI values from these motion encrypted ECG signals is critical. Even if IBI is directly estimated under this heavy motion situation, it will be nowhere near perfect. So, there is an unmet need for noise rejection techniques that can deal with high level of noise and motion artifact. 

To overcome aforementioned unmet need, we propose a novel framework to determine IBI values from ECG signals totally buried in intense motion artifact noise. We have modified \textit{The One Hundred Layers Tiramisu} network \cite{Jgou29}, made it suitable for time-series denoising process, and used it for our case. The intention of this study is to eliminate motion artifact and make the R-peaks of an ECG signal more prominent in a way such that IBI values can be estimated from it as accurately as possible. Our modified network works as an autoencoder. An autoencoder is a well-known framework for making compact representation of a signal and taking it back to its original dimensions if required. Our methodology intends to squeeze the given ECG signal using autoencoder, suppress the noise, get back to the original dimensions, and represent the ECG signal with its peaks sufficiently visible to calculate IBI values. The autoencoder we have used is comprised of fully convolutional dense networks (FC DenseNet) - a Tiramisu model - which is familiar for image segmentation, image classification and more recently time domain analysis.

In the process of estimating IBI values from noisy ECGs, the contribution of this paper can be highlighted as follows.
\begin{itemize}
    \item[1.] An FC DenseNet based deep learning approach - derived from \textit{The One Hundred Layers Tiramisu model} \cite{Jgou29} - which performs as a robust autoencoder to suppress the noise and makes the R-peaks of the noisy ECG signal more notable. To the best of our knowledge, such a stacked tiramisu model is yet to be applied in beat detection from ECG with high intensity noise and IBI estimation as well.
    
    \item[2.] Our proposed methodology is capable of reckoning IBI values from ECG with noise level up to -30dB with some considerable amount of error. So far, we haven't noticed other works going beyond this range without the inclusion of high frequency or Gaussian white noise.
    
    \item[3.] In contrast to many concurrent works, our proposed approach does not require any sort of pre-processing or post-processing algorithms or tools. However, a simple peak picking algorithm has been used to facilitate the IBI calculation procedure.
    
\end{itemize}

The remaining of this paper is organized as follows. Related works are discussed in Section \ref{rltdwrks}. The dataset used in this work, noise signals, their characteristics, and noise addition protocols are discussed in Section \ref{dtst}, development of the proposed method is explained in Section \ref{prpsdmthd}, results, detailed comparison with state of the art work in Section \ref{rslts}, observations, drawbacks, and discussions are presented in Section \ref{disc}. In the end, we conclude with a brief conclusion in Section \ref{conc}.

\section{Related work}
\label{rltdwrks}

Research in the field of IBI estimation and ECG R-peak detection, which is the requirement of IBI estimation, from noisy ECG is high in number. Researchers have used a plethora of signal processing tools and techniques for accurate estimation of IBI in presence of motion artifacts. Mostly, they have focused on some publicly available ECG datasets like MIT-BIH Arrhythmia database or IEEE SP Cup Dataset (2015).

Removing high frequency noise or Gaussian noise from ECG and then estimating IBI from it can be achieved by applying  straightforward moving average \cite{Pandey09}, low-pass, high-pass or band-pass FIR filters \cite{Almal10,Rakshit11}, Empirical Mode Decomposition (EMD) technique \cite{Karagiannis12,Talbi13}, Pan-Tompkins peak detection \cite{Pan1985ARQ}, and wavelet transformation \cite{Sundararaj14}. However, the challenge starts when we tend to do the same in presence of heavy motion artifacts. Again, baseline wander and power line noise can be removed by using band-pass filter and notch filter. But, electrode motion artifact noise - which is caused by skin stretching and alters the impedance of the skin around the electrode - takes rigorous process to be encountered as its frequency spectrum (1 to 10 Hz) overlaps with that of the PQRST complex of ECG waveform \cite{Kher15}.

\textit{Rahel et al.} designed a study to evaluate the IBI signal qualities of a Holter device and a heart rate chest belt monitor while the subjects were at rest and were performing 5 different levels of activities like sitting and reading, doing household chores, walking, jogging, and training \cite{GilgenAmmann16}. But there is no indication of the specific amount of noise they dealt with. \textit{Shalini et al.} used Slope Sum Function and Teager-Kaiser Energy operator method for R-peak artifacts detection in cardiovascular and non-cardiovascular signals like Electroencephalogram (EEG), Electrooculogram (EOG), and Electromyogram (EMG) \cite{Rankawat22}. However, they too didn't mention the SNR level up to which their algorithm works perfectly. 

\begin{figure*}[t]
\centering
\includegraphics[width=0.7\textwidth]{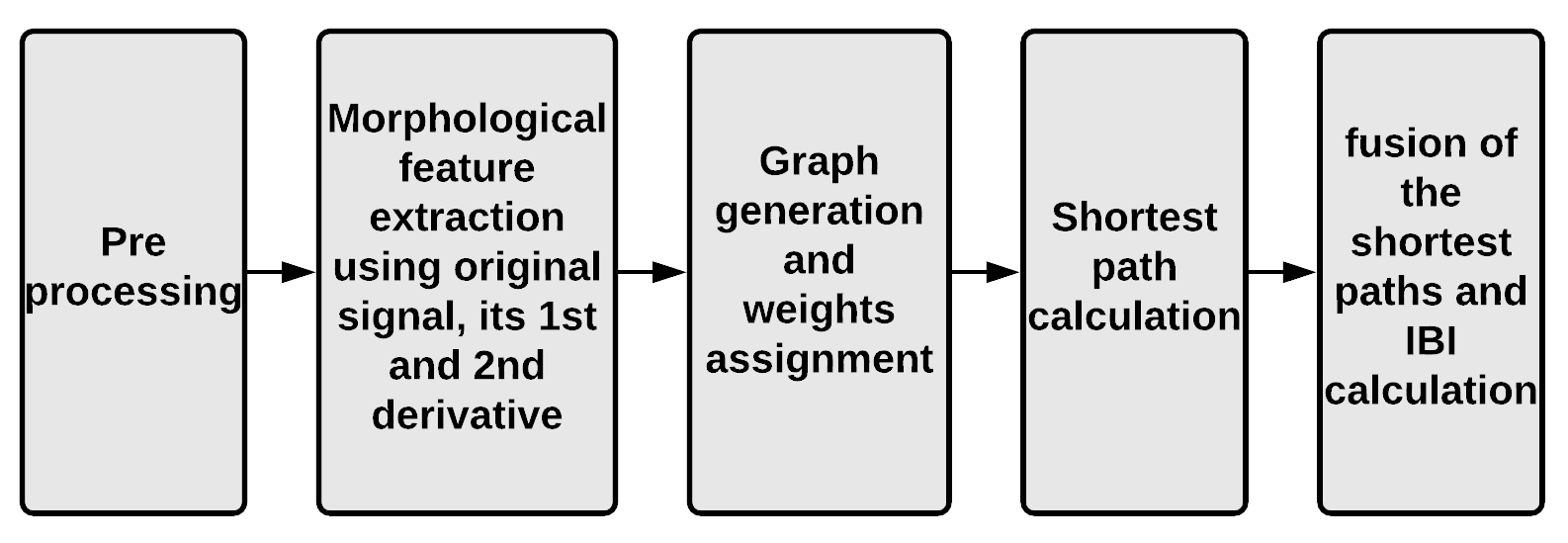}
\caption{Overview of the work performed by \textit{Aygun et al.} \cite{Aygun18}.} 
\label{fig:aycya_flow}
\end{figure*}

\textit{Sonia et al.} discussed a new time delay estimation technique which is essentially derived from operational calculus, differential algebra, and non commutative algebra and helps to estimate RR interval from noisy ECG with SNR level up to 6dB with considerable number of false peak detection \cite{Rezk19}. \textit{Aygun et al.} proposed a technique to select the fiducial points from noisy ECG and PPG signal using shortest path algorithm which takes the advantage of time-continuity of heartbeats \cite{Aygun17}. With this, they accurately measured IBI up to -2dB SNR level which they later used for determining HRV \cite{Aygun18}. Here, the derivatives of the ECG or PPG signal have calculated using Savitzky-Golay method to detect the possible fiducial points like R-peaks, systolic peaks, points with maximum slope, and onset points. Since the signals are noisy, they ended up getting too many candidate points and most of them were false fiducial points if not all. With all these points, a graph was generated for each signal where the edges refer to IBIs, both true and false IBIs, and the vertices refer to morphological points, also both true and false. From this graph, it was clear that the starting point of one heartbeat is the end point of the last one and there is no disruption between them. Any candidate path that satisfies certain conditions derived from average HR can be considered as a true IBI and the two associated nodes/vertices as true morphological point for that specific signal. Next, each of the vertices were connected to some previous vertices which fell within a specific time window. If the time difference between the reference vertex and any neighboring vertex deviated the average IBI (which was calculated using average HR), the corresponding edge i.e. the 'time interval' was assigned zero weight. Otherwise, a numerical weight was assigned to it where the weight varied according to its difference with the average IBI. The intuition behind this weighting is the fact that the true IBIs should remain close to the average IBI calculated from the average HR. The weights of the vertices were also derived from the weights of corresponding edges. After all these weight distribution, the vertex with the minimum accumulated weight, is the chosen true vertex inside the time window. This is indeed leveraging the shortest path algorithm and enables detecting the IBIs that are closest to the average IBI. Finally, they combined all the shortest paths and formed an array for each of the morphological features of each ECG and PPG signal. An overview of this process is depicted in Figure~\ref{fig:aycya_flow}. Although both \cite{Aygun18} and \cite{Rezk19} have employed novel mathematical tools for IBI estimation without necessarily denoising the entire ECG signal, they haven't been able to go far in terms of SNR level.

Many researchers have focused on R-peak detection from noisy ECG signals. As mentioned earlier, a robust peak detection algorithm can also lead to accurate IBI estimation. In addition to applying a band-pass filter for removing power line and baseline shifting noise, \textit{Pulavskyi et al.} performed a two-stage smoothing using 'boxcar' and 'parzen' kernel. This methodology allowed them to detect peaks with good precision and recall values up to -15dB, however, there was presence of white and pink noise which decreased the relative weight of electrode motion and muscle artifacts  \cite{Pulavskyi20}. 

\begin{figure*}[h]
\centering
\includegraphics[width=0.8\textwidth]{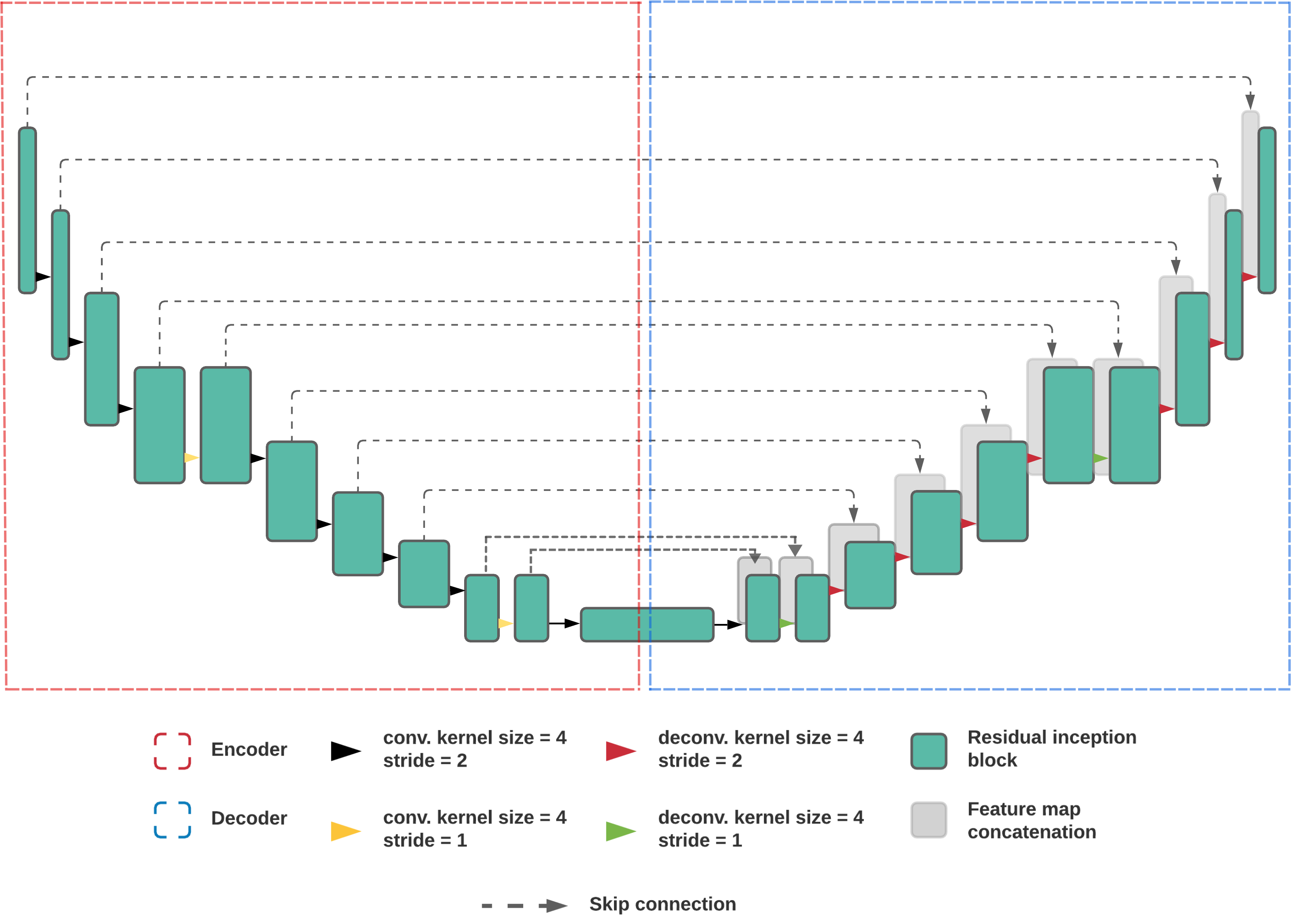}
\caption{The U-Net architecture used by \textit{Sricharan et al.} \cite{Vijayarangan26}.} 
\label{fig:Sri}
\end{figure*}

With advanced deep learning mechanisms, denoising tasks have gone further as they have been able to obtain higher accuracy and hardly require human supervision. To talk about a few deep learning based approaches, \textit{Ansari et al.} exploited a simple convolutional neural network (CNN) to differentiate between usable and unusable ECG segments. The usable segments have higher probability of QRS detection. They carried out their research up to 0dB with some error \cite{Ansari21}. \textit{Juho et al.}, on the other hand, utilized a bidirectional Long-Short Term Memory (LSTM) network to suppress the noise up to 0.1dB and detect peaks from it \cite{Laitala23}. A CNN encoder-decoder is used by \textit{Natasa et al.} to denoise the QRS complexes of the long term ECG signals acquired with their wearable armband device and used them later to calculate HR \cite{Reljin24}. They too were able to use the denoising process up to -17dB with inclusion of white and brown noise, but no muscle artifact. \textit{Sricharan et al.} modified the conventional U-Net \cite{Ronneberger25} algorithm for time-series data and used it along with distance transformation to determine the position of the ECG R-peaks for noise level up to 0dB \cite{Vijayarangan26}. The peak detection task was framed as a regression task in \cite{Vijayarangan26}. They obtained the Distance Transformation (DT) of all the ECG signals which disclose the distance of each point on the ECG from its nearest peak. It results in a zig-zag wave where the lowest points refer to the R-peaks of the corresponding ECG signal. The DT is of the same size as its input ECG. They have exploited the U-Net architecture with conventional encoder-decoder where the encoder performs downsampling with 8 layers of strided-convolutions, the decoder does upsampling in a similar way but in opposite direction. Both the encoder and decoder perform 1D convolution and there is a bottleneck between them to hold the minimum representation. Additionally, a residual inception block was placed at each layer which exploits skip connections to discard vanishing gradient problem and ensure fast convergence. The architecture for this work is illustrated in Figure~\ref{fig:Sri}. 

\begin{figure*}[h]
\centering
\includegraphics[width=0.9\textwidth]{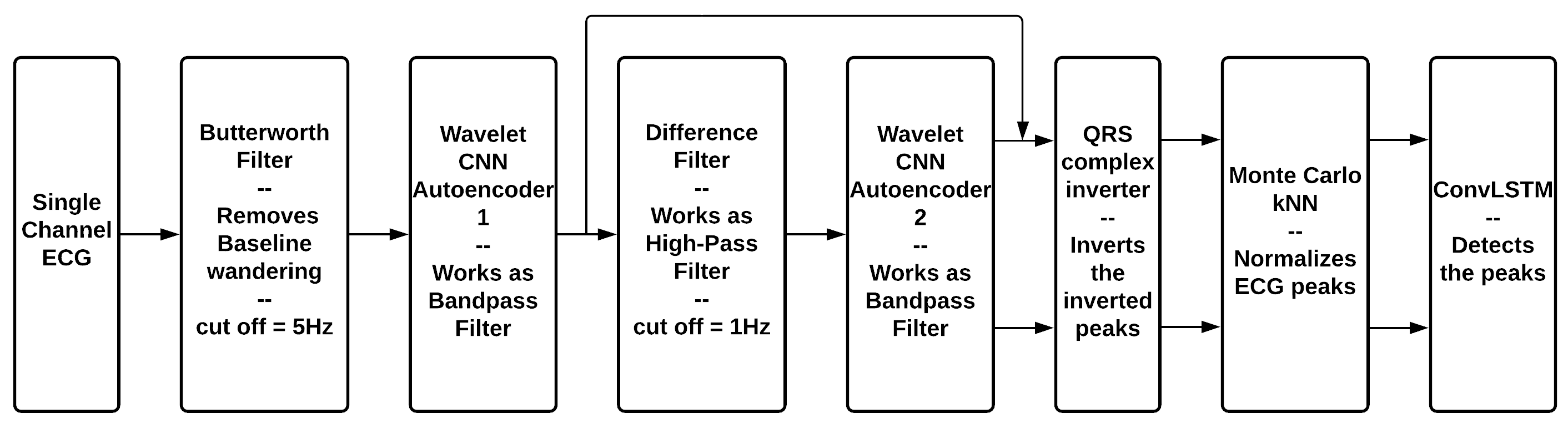}
\caption{The machine learning pipeline developed by \textit{Brosnan et al.} \cite{Yuen28}.} 
\label{fig:pipeline}
\end{figure*}

Likely, \textit{Lishen et al.} took the benefit of U-Net and DR-Net to perform a two-stage denoising process successfully \cite{Qiu27}. A generative adversarial network (GAN) based ECG synthesizer has been designed by \textit{Karol} to generate his own synthesized ECG dataset. Later, noise was added to the artificial ECG and a CNN autoencoder was trained  utilizing this dataset to achieve an MSE of 0.017 (direct from optimizer during training) \cite{Antczak2020AGA}. Finally, \textit{Brosnan et al.} \cite{Yuen28} demonstrated peak detection with a machine learning pipeline that consists of a Butterworth filter, two wavelet convolutional neural networks (WaveletCNNs) autoencoders, an optional QRS complex inverter, a Monte Carlo k-nearest neighbours (k-NN), and a convolutional long short-term memory (ConvLSTM) as shown on Figure~\ref{fig:pipeline}.

The noisy one-channel ECG signal is fed into the Butterworth filter with cut-off frequency set at 5Hz to perform the task of removing baseline shifting noise by attenuating the low frequency components. Next, the output is decomposed using symlet-4 wavelet to get the wavelet coefficients. The coefficients are fed into a CNN autoencoder where the noisy unnecessary coefficients are rejected but the significant parts retain. Thus it produces clean wavelet coefficients which are significant for further analysis. The CNN autoencoder that takes on the wavelet coefficients has a 2 layered encoder, a bottleneck and a 2 layered decoder and Leaky-ReLU activation layer at each hidden layer. A wavelet reconstruction is performed to get the coefficients back to original form i.e. single channel ECG. Now, a difference filter is deployed prior to the next wavelet autoencoder to refine the QRS complexes by suppressing the low frequency components. Next, the signal is moved to the second wavelet autoencoder which has same functionality as the previous one. Both of the two wavelet autoencoder work as bandpass filters to remove the unnecessary components. To deal with the inverted peaks, an inverter was placed right after the second autoencoder and it just inverted the already inverted peaks. Then Monte Carlo k-NN method was utilized to normalize the ECGs by scaling the beats to unit mili-volt. Finally, the ConvLSTM network was applied to detect the peak points as it has pattern matching ability to detect any specific characteristics from a time-series signal.They used the MIT-BIH, the European ST-T and the Long Term ST database Noise Stress Test databases, went up to -6dB and recorded some good result. All these deep learning based methods, however, failed to leverage different deep learning algorithms' power to improve the accuracy enough for denoising or IBI estimation and explore intense noise scenarios.

\section{Datasets}
\label{dtst}
We have opted to use clean ECG recordings from publicly available database like MIT-BIH Arrhythmia Database v1.0.0 \cite{Moody2001TheIO,Goldberger2000PhysioBankPA}, European ST-T Database from PhysioNet \cite{Taddei1992TheES} and IEEE Signal Processing Cup 2015 \cite{Zhang2015TROIKAAG}. 

The sampling frequency of MIT-BIH Arrhythmia Database is 360 Hz i.e. 1 sample in every 2.78 ms. For ease of calculation and noise addition, we upsampled both European ST-T Database and IEEE SP Cup 2015 dataset from 250 Hz and 125 Hz to 360 Hz respectively. 

We then added electrode motion (EM) noise, motion artifact (MA) noise and baseline wandering (BW) noise from  MIT-BIH Noise Stress Test (NST) Database v1.0.0 \cite{Moody1992TheMN} and generated noisy data for different SNR levels starting from 36dB to -36dB with a decrement of 6dB. So, for each clean ECG signal, we generated 13 variants with different SNR levels. Needless to mention, noise have been added to the signals thoroughly instead of adding them between consecutive rest periods. For this noise addition task we followed the rules given in NST Database-
\begin{equation*}
\textit{ECG}\textsubscript{noisy} = \textit{ECG}\textsubscript{clean} + [a_1]*\textit{EM} + [a_2]*\textit{MA} + [a_3]*\textit{BW}
\end{equation*}

where, $\textit{ECG}\textsubscript{clean}$ refers to clean ECG signal and $a_1, a_2, a_3$ gain values are adjusted according to the expected SNR value. Since we are more interested in estimating IBI values from a signal totally buried under muscle artifact and electrode motion, these two noises have been assigned higher weights meaning $a_2 > a_1 > a_3$.

\begin{figure}[!htbp]
\centering
\includegraphics[scale = 0.28]{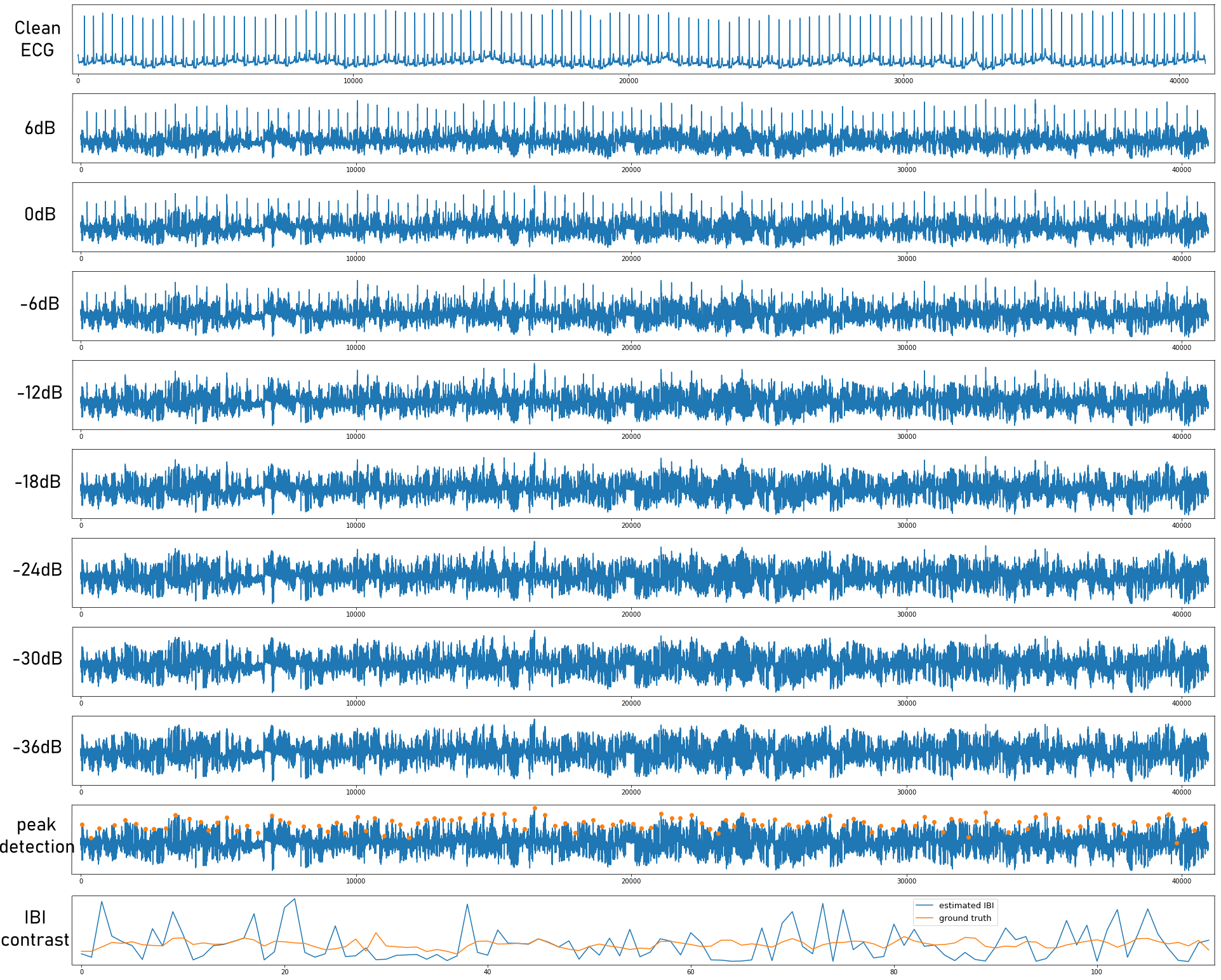}
\caption{A clean ECG segment and its noisy variants. EM, MA, BW noises have been added to clean signals and the numbers on the left of each box refer to corresponding SNR values. IBIs estimated from raw and noisy ECG signal are highly inaccurate compared to the true IBIs} 
\label{fig:noise}
\end{figure}

\begin{table*}[b]
\begin{center}
\caption{List of used ECG recordings from different database}
\setlength{\arrayrulewidth}{0.4mm}
\setlength{\tabcolsep}{18pt}
\resizebox{0.7\textwidth}{!}{%
\begin{tabular}{|cc|c|c|}
\hline
\multicolumn{2}{|c|}{MIT-BIH Arrhythmia Database} & 
European ST-T Database & 
IEEE SP Cup 2015 \\
\hline
\multicolumn{2}{|c|}{Train data} & Train data & Train data \\ \hline
100 & 102 & e0103 & 1 \\
 & & e0104 & 5 \\ \hline
\multicolumn{2}{|c|}{Test data} & Test data & Test data \\ \hline
101 & 209 & e0105 & 2 \\
103 & 213 & e0106 & 3 \\
109 & 219 & e0107 & 4 \\
112 & 220 & e0113 & 6 \\
113 & 223 & e0114 & 7 \\
115 & 228 & e0115 & 8 \\
116 & 230 & e0118 & 9 \\
122 & 231 & e0122 & 10 \\
123 & 234 & e0123 & 11 \\
&  & e0126 & 12 \\
&  & e0127 &  \\
&  & e0129 &  \\

\hline
\end{tabular}}
\label{tab1}
\end{center}
\end{table*}

In this work, we used first 409600 data points of each ECG signal since the noise signals have limited data points. Figure~\ref{fig:noise} shows a clean segment of ECG data from MIT-BIH Arrhythmia database (recording \# 101) and its corresponding noisy variants starting from 6db to -36dB. As the noise increases, specifically after 0 dB, the R-peaks start to heavily get buried in noise and detecting them with conventional signal processing techniques would be very inaccurate and challenging if even possible. These artifacts enforce redundant tests, additional costs, and specialists' intervention sometimes. Table~\ref{tab1} lists the recordings from different dataset used for training and testing. Usually, autoencoder type models require very small training data \cite{Vijayarangan26} and our case is no different.

\section{Proposed Methodology}
\label{prpsdmthd}

\begin{figure*}[!htbp]
\centering
\includegraphics[width=0.9\textwidth]{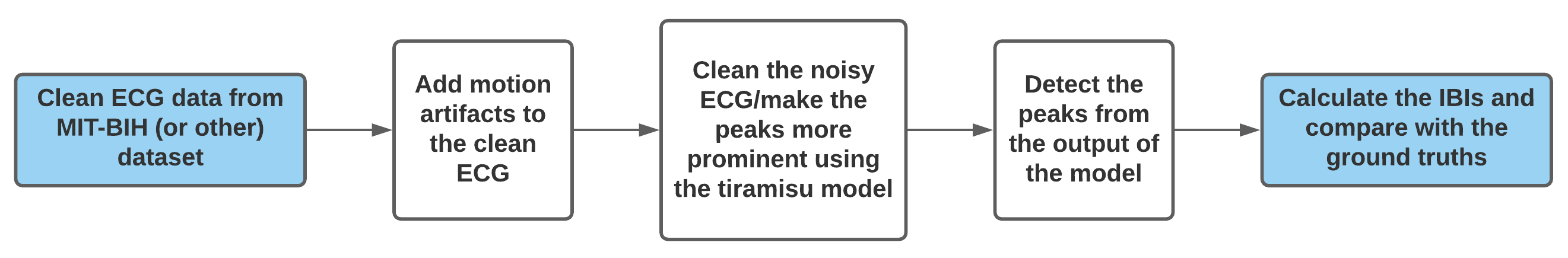}
\caption{An overview of the proposed IBI estimation framework.} 
\label{fig:flow}
\end{figure*}

Our working principle starts with a fully convolutional-dense deep neural network (tiramisu model) \cite{Jgou29}. Noisy signals are fed to this model which has multiple convolution and pooling layers intuitively responsible for compressing the signal to a smaller representation, eliminating the noise, retaining the most patterned and prominent features, and making the peaks more visible. Later, we identify the peaks with a simple peak picking algorithm and then calculate IBI values from there. The overall procedure of this work is illustrated in Figure~\ref{fig:flow}. 

\subsection{Tiramisu Model}
It is challenging to calculate IBI values from noisy ECG signals. As a solution, we designed a robust tiramisu model which consists of stacked dense blocks (DB) and numerous feedback loops. High number of dense blocks and feedback loops eliminate the requirement of pre-processing as the model itself is capable of removing high frequency or Gaussian white noise. While the convolution and pooling layers perform the job of squeezing out the noise, the added skip connections between layers allow the output of each dense block and up convolution to encode finer details from actual signal and later layers, solve the gradient vanishing problem, and provide improved results than without them. In contrast to the standard U-Net architecture, the dense blocks in the tiramisu model have skip connections within themselves which is a fact that eliminates the possibility of gradient vanishing within the contraction path. 

\begin{figure}[t]
\centering
\includegraphics[scale=0.18]{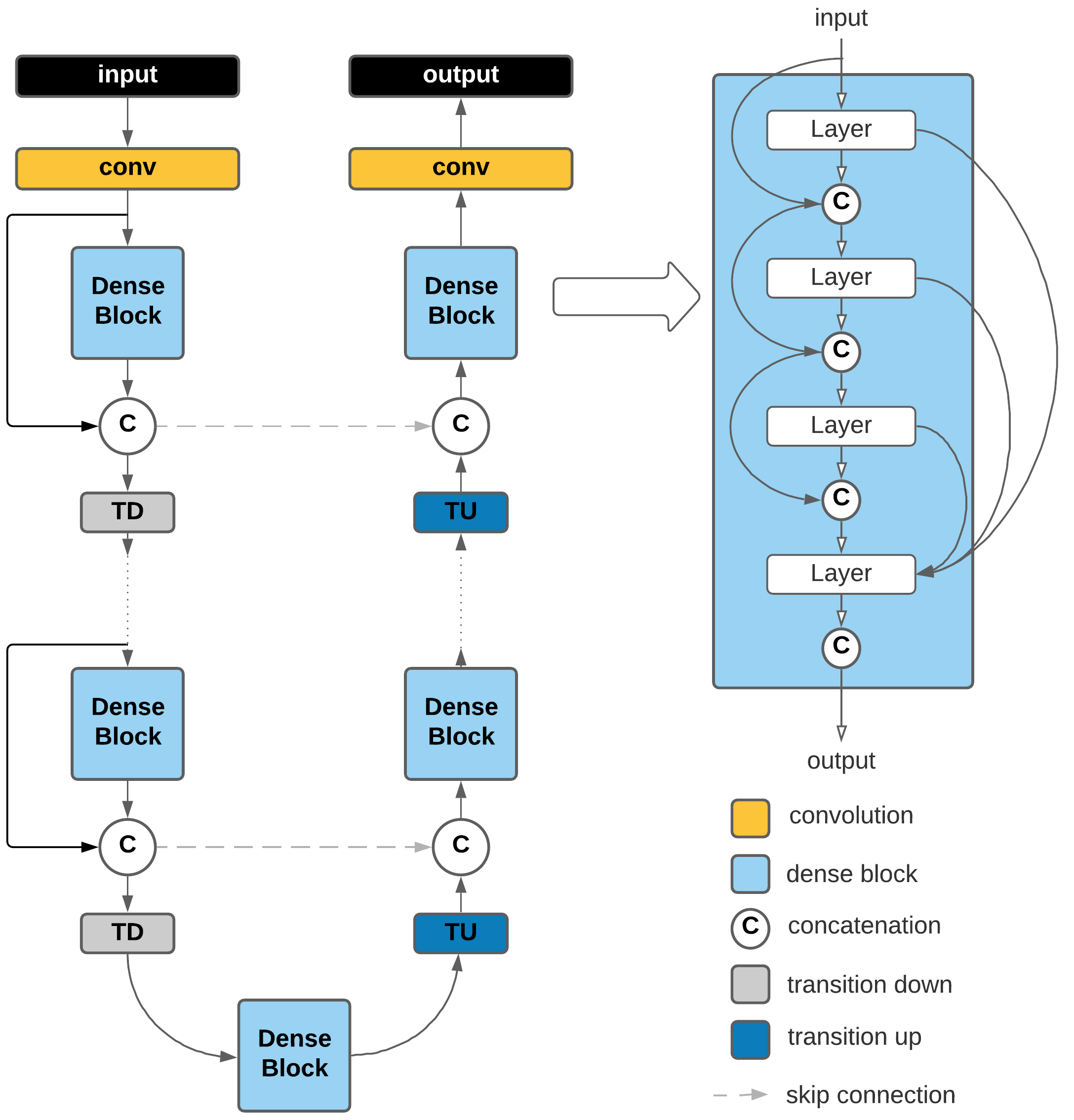}
\caption{Big picture view of the proposed model. It consists of a contraction and an expanding path. Both paths have several convolution blocks, dense blocks, transition downs, transition ups, concatenations and skip connections. The dense block between these two paths is the bottleneck.} 
\label{fig:tiramisu}
\end{figure}

The overall architecture of the proposed model is illustrated in Figure~\ref{fig:tiramisu}. As depicted, the model incorporates a contraction path, a bottleneck, an expanding path, and certain skip connections. The skip connections here help the expanding path (or upsampling path) to get back information from the contraction path, reuse them and thus abolish gradient vanishing. The main theme of this entire model is to take advantage of feature reuse by taking the already complicated DenseNet a step ahead. 

The contraction path (or downsampling path) starts with a convolution with kernel size = 3. Then comes a series of consecutive dense blocks and transition down (TD) blocks. Prior to the bottleneck, 3 dense blocks have been placed on the contraction path with each having 4, 5, and 7 layers respectively. Right after each dense block, a TD block has been attached. Note that, the TD blocks perform the pooling operation and cause some information loss and resolution reduction along the downsampling path. In between, there are concatenation blocks to create feedback loops from previous layers.

The expanding path is almost a mirror symmetry of the contraction path except it has lesser feedback loops and the TD blocks are replaced by transition up (TU) blocks along the path. These TU blocks does transposed convolution operation and upsample the previous mappings. The upsampled mappings are then concatenated to the mappings coming from the downsampling path via skip connections and form a new input for the next dense blocks along expanding path. The last dense block in this trajectory sums up the information from all the previous dense blocks. Unlike the contraction path, we do not concatenate the input of a dense block with its output since it does not make any significant difference. The final convolution in this path is done with widow size = 3. The dense block that lies between these two paths is the bottleneck.

The dense blocks in the model have different number of layers which perform one dimensional convolution with kernel size = 3, \textit{ReLU} activation function, Batch Normalization, and drop out value of 0.2. The batch normalization function is given by-

\begin{equation}
    BN(x) = \frac{x-\mu_x}{\sigma_x}
\end{equation}

where, $\mu_x$ is the mean and $\sigma_x$ is the standard deviation of input matrix $x$.

The TD layers also use one dimensional convolution with kernel size = 1, \textit{ReLU} activation, Batch Normalization, drop out value of 0.2 and Max Pooling with pool size = 2. However, the TU layers perform transposed convolution with kernel size = 3 and stride = 2.

On the final output layer we have employed \textit{tanh} activation since we desire to have our output between [-1 1] and gain non-linearity for getting advantage during derivative calculation. The \textit{tanh} function can be written as-

\begin{equation}
    tanh(x) = \frac{e^{x}-e^{-x}}{e^{x}+e^{-x}}
\end{equation}

As mentioned earlier, we have 3 dense blocks on either path with each block having 4, 5, and 7 layers respectively. Also, we have 3 TUs or TDs and 1 input or output block on either side. Since each layer has one convolution layer, each TU or TD also perform one convolution and each input/output block has one convolution layer, we have 2*(4+5+7+3+1) = 40 convolution layers on these two paths combined. Additionally, the bottle neck dense block has 10 layers i.e. 10 convolution layers. So, a total of 50 convolutions are done in this entire model. All the convolutions (except last one) are equipped with \textit{ReLU} activation, L2 regularization with factor of 0.01 and zero padding with a view to redeeming the initial shape. Finally, we have employed the Adam optimizer with default learning rate i.e. 0.001 and mean square error (MSE) based loss function-
\begin{equation}
    MSE = \frac{1}{N}\sum_{i=1}^{N}(y_i - y_i^p)^2
\end{equation}

where, $y_i$ and $y_i^p$ refer to actual target value and predicted value respectively and $N$ is the number of data points.

Table \ref{tab2} shows the full architecture of our tiramisu model and the constituents of layer, TD, and TU blocks.

\begin{table}[!htbp]
    \caption{Full tiramisu model architecture and contents of layer, TD and TU blocks.}
    \begin{minipage}{.32\linewidth}
      \centering
        \begin{tabular}{||p{4cm}||}
        \hline
           \begin{center}\textbf{Full tiramisu model architecture}\end{center}\\
         \hline\hline  
         Input\\
         \hline
         Convolution: kernel size = 3\\
         \hline
         DB (4 layers) + TD \\
         \hline
         DB (5 layers) + TD\\
         \hline
         DB (7 layers) + TD\\
         \hline
         Bottleneck: DB (10 layers)\\
         \hline
         DB (7 layers) + TU\\
         \hline
         DB (5 layers) + TU\\
         \hline
         DB (4 layers) + TU\\
         \hline
         Convolution: kernel size = 3\\
         \hline
         \textit{tanh} activation\\
         \hline
         Output\\
         \hline
        \end{tabular}
    \end{minipage}%
    \begin{minipage}{.4\linewidth}
      \centering
        \begin{tabular}{||p{3.2cm}||p{3.2cm}||p{3cm}||}
        \hline
            \begin{center}\textbf{Layer block}\end{center} & \begin{center}\textbf{Transition Down (TD) block}\end{center}&\begin{center}\textbf{Transition Up (TU) block}\end{center}\\
        \hline\hline
        Batch Normalization & Batch Normalization & Trans. convolution: kernel size = 3\\
        \hline
        \textit{ReLU} activation & \textit{ReLU} activation & l2 regularizer (0.01)\\
        \hline
        l2 regularizer (0.01) & l2 regularizer (0.01) & \textit{stride} = 2\\
        \hline
        Convolution: kernel size = 3 & Convolution: kernel size = 1 &\\
        \hline
        Dropout $p$ = 0.2 & Dropout $p$ = 0.2 &\\
        \hline
        &Max Pooling: size = 2&\\
        \hline
        \end{tabular}
        \label{tab2}
    \end{minipage} 
\end{table}

\subsection{Peak Detection}
For peak detection, we have employed a general algorithm \cite{peak}. Usually, resting HR varies from 60-100 bpm \cite{American_Health}. But for athletes, it could be as low as 40 bpm during rest periods and as high as 200 bpm during work out sessions \cite{Athletes}. Our sampling frequency is 360 Hz. It ensures that there must be at least one beat in every 108 to 540 data points. So we have used an algorithm that takes on a given signal and finds out all the local maxima which must be at least 108 data points apart from each other. In this way, we may end up getting some false peaks because, when the noise is intense and a motion artifact-induced peak is higher than its neighboring R-peak, the tiramisu model fails to detect the R-peak and removes it while the motion artifact remains. Subsequently, the peak detection algorithm detects this motion artifact as an R-peak causing some error.

\subsection{IBI Calculation}
Under any circumstance, to estimate IBI precisely, the position of the beats/peaks need to be detected accurately. After denoising the signal using the proposed AE-based model and detecting the R-peaks, in this step, the IBI value using the identified R-peaks can be extracted with Equation \ref{eq1}.
\begin{equation} \label{eq1}
    IBI_{i} = t_i-t_{i-1}
\end{equation}

\noindent where, $t_i$ is the occurrence time of the $i^{th}$ R-peak and IBI is extracted for all peaks in an ECG signal. 

\section{Results}
\label{rslts}

In this study, we aim to make the R-peaks of noisy ECG more prominent to estimate IBI from it as accurately as possible. We evaluate the performance of our approach by testing it on different levels of noisy ECG, which have been generated by adding certain level of noise, explained in Section \ref{dtst}. In this section, we define our evaluation metrics and measure the performance of our model in terms of estimated IBI values extracted from ECGs with different SNR levels. Lastly, we will compare our approach against a few state-of-the-art methods designed for IBI calculation and/or peak detection from noisy ECG.

\begin{figure*}[t]
\centering
\includegraphics[width=\textwidth, height=9cm]{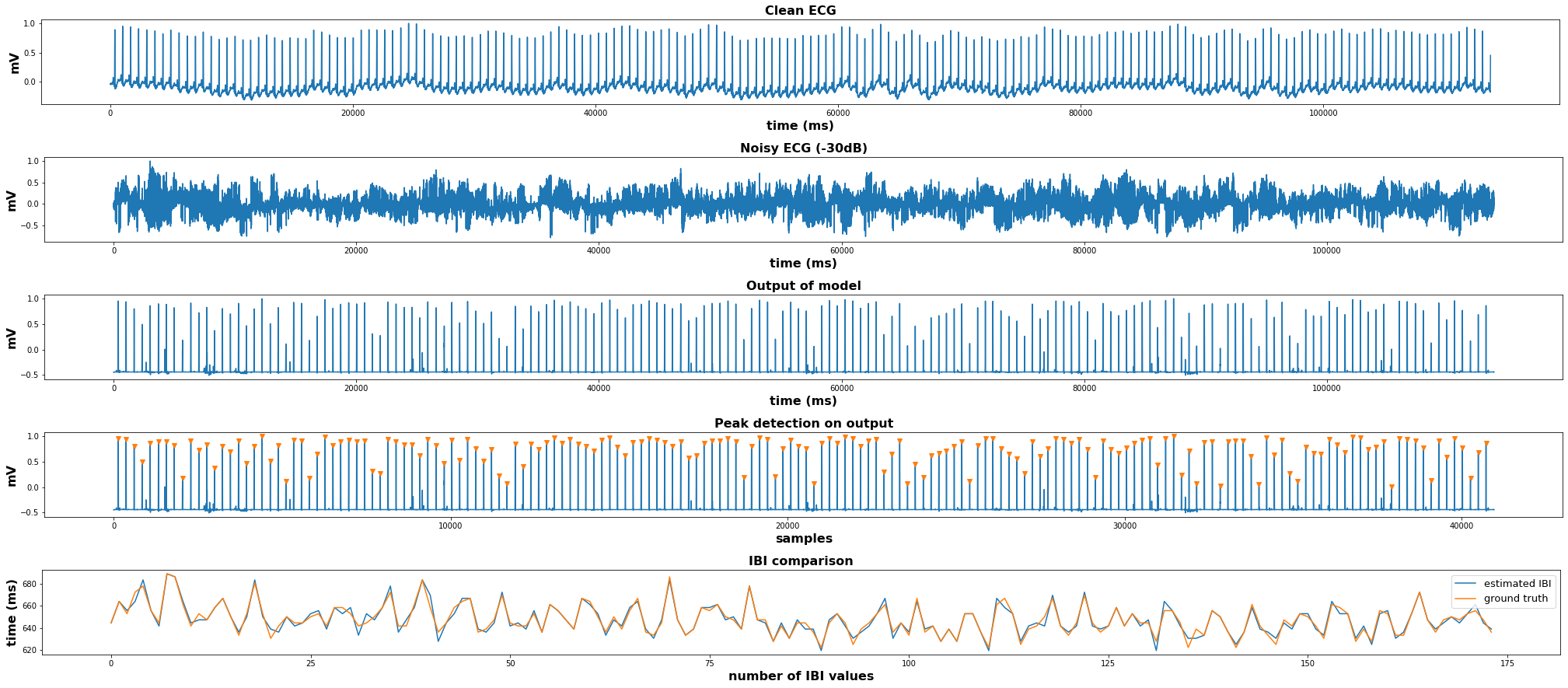}
\caption{\textbf{\textit{Clean ECG:}} An ECG segment from MIT-BIH Arrhythmia Database (recording \# 234). \textbf{\textit{Noisy ECG:}} Muscle artifact, baseline wander and electrode motion noise have been added to the clean signal. Resulting SNR of the ECG is -30dB. \textbf{\textit{Model output:}} The noisy ECG is fed to the tiramisu model and the output has the R-peaks more prominent. \textbf{\textit{Peak Detection:}} Peak detection algorithm has been applied to the output. \textbf{\textit{IBI Comparison:}} IBI values estimated from the prominent peaks and compared against the ground truth IBI values. The RMSE of IBI values in this case is only 3.44 ms.} 
\label{fig:demo}
\end{figure*}

For further explanation we refer to Figure~\ref{fig:demo}. The clean ECG in this figure is the first 40960 data points of recording 234 from MIT-BIH Arrhythmia Database. Three different noises have been added to the clean ECG to get a noisy signal of -30dB SNR. -30dB refers to a very high amount of noise and implies that the power of noise is $10^3$ times that of actual signal.  Without any pre-processing, this noisy ECG is then fed to the tiramisu model to make the R-peaks more visible. Next, our peak detection algorithm has been applied to the output of the model. We estimated IBI values from the beats and in the final box, we have compared our IBI values with that of the ground truth and recorded a root mean squared error of 3.44 ms.

\begin{figure}[t]
     \centering
     \begin{minipage}[!htbp]{0.24\textwidth}
         \centering
         \includegraphics[width=\textwidth]{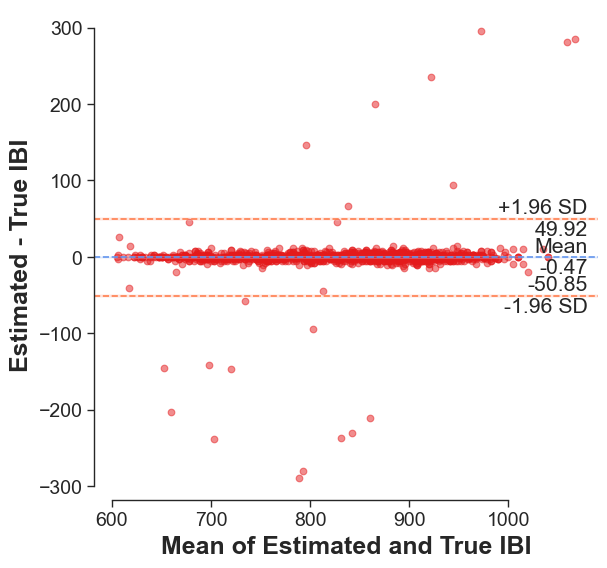}\\
         \subcaption{recording \# 230}
         \label{BA1}
     \end{minipage}%
     \hfill
     \begin{minipage}[!htbp]{0.24\textwidth}
         \centering
         \includegraphics[width=\textwidth]{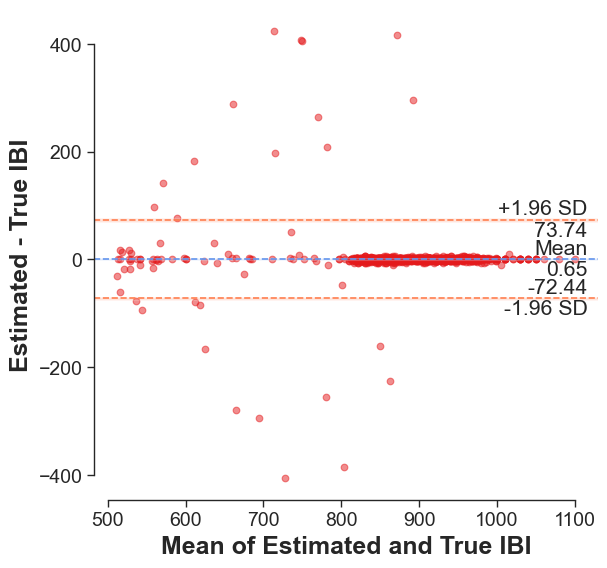}\\
         \subcaption{recording \# 220}
         \label{BA2}
     \end{minipage}%
     \hfill
     \begin{minipage}[!htbp]{0.24\textwidth}
         \centering
         \includegraphics[width=\textwidth]{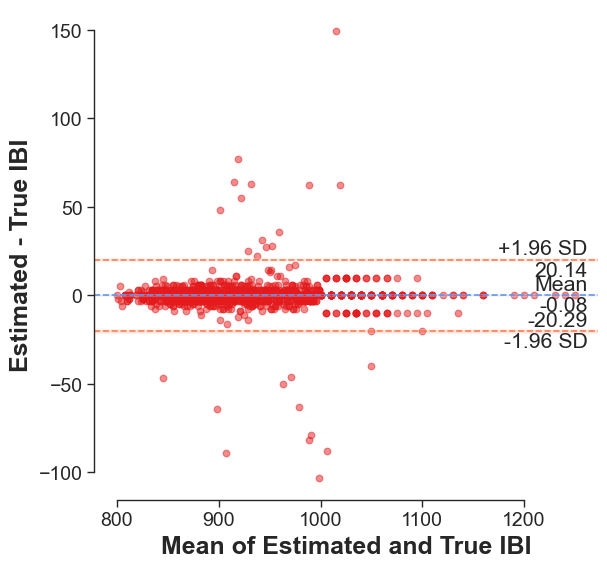}
         \subcaption{recording \# 115}
         \label{BA3}
     \end{minipage}%
     \hfill
     \begin{minipage}[!htbp]{0.24\textwidth}
         \centering
         \includegraphics[width=\textwidth]{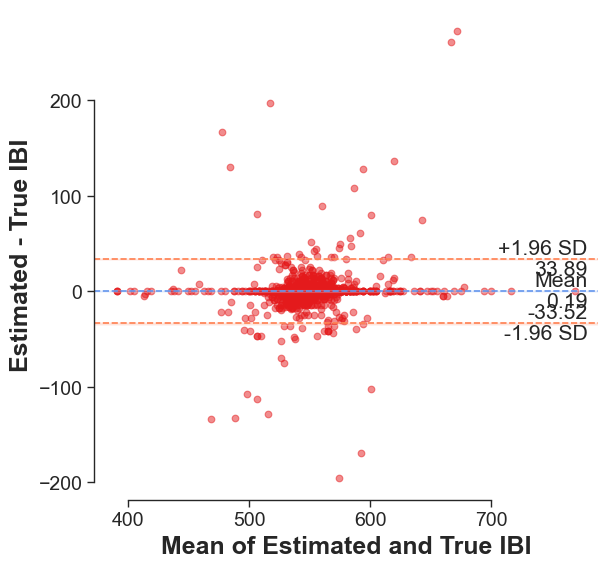}
         \subcaption{recording \# 213}
         \label{BA4}
     \end{minipage}%

     \begin{minipage}[!htbp]{0.24\textwidth}
         \centering
         \includegraphics[width=\textwidth]{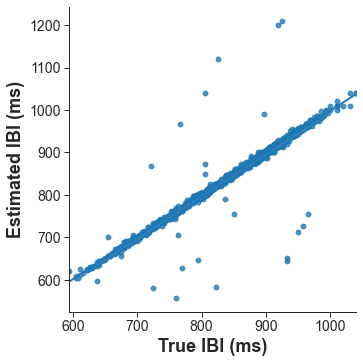}
         \subcaption{recording \# 230\\\textit{r} = 0.9595}
         \label{C1}
     \end{minipage}%
     \hfill
     \begin{minipage}[!htbp]{0.24\textwidth}
         \centering
         \includegraphics[width=\textwidth]{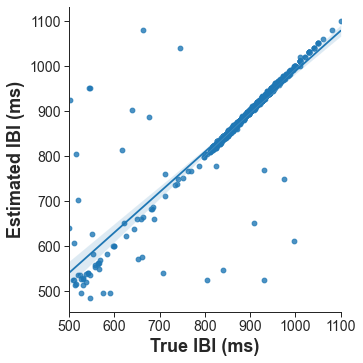}
         \subcaption{recording \# 220\\\textit{r} = 0.9273}
         \label{C2}
     \end{minipage}%
     \hfill
     \begin{minipage}[!htbp]{0.24\textwidth}
         \centering
         \includegraphics[width=\textwidth]{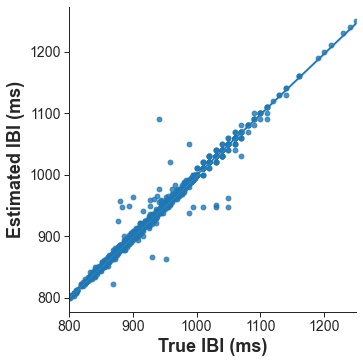}
         \subcaption{recording \# 115\\\textit{r} = 0.9895}
         \label{C3}
     \end{minipage}%
     \hfill
     \begin{minipage}[!htbp]{0.24\textwidth}
         \centering
         \includegraphics[width=\textwidth]{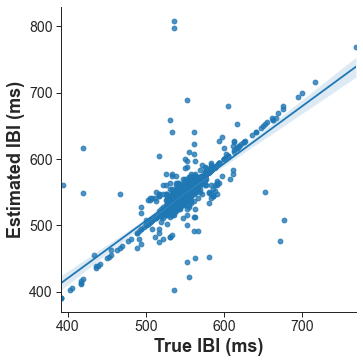}
         \subcaption{recording \# 213\\\textit{r} = 0.9006}
         \label{C4}
     \end{minipage}%
     
        \caption{Bland-Altman plot with LOAs are depicted in Figs. \ref{BA1} to \ref{BA4} for different recordings. Also, correlation plot for same recordings are shown with Pearson correlation coefficient in Figs. \ref{C1} to \ref{C4}. For some ECGs, the proposed method achieves r values like 0.9895 though nominal values like 0.9 also exist.}
        \label{fig:seven graphs}
\end{figure}

\begin{figure}[!htbp]
     \centering
     \begin{minipage}[!htbp]{0.25\textwidth}
         \centering
         \includegraphics[width=\textwidth]{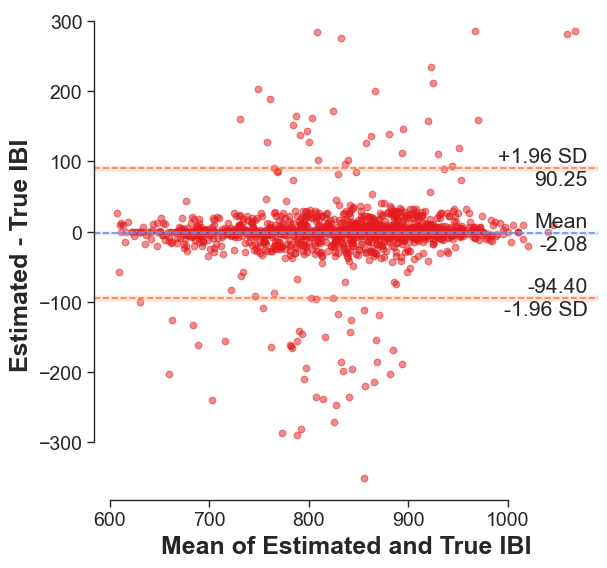}\\
         \subcaption{\textit{Brosnan et al.}}
         \label{BA5}
     \end{minipage}%
     \hfill
     \begin{minipage}[!htbp]{0.25\textwidth}
         \centering
         \includegraphics[width=\textwidth]{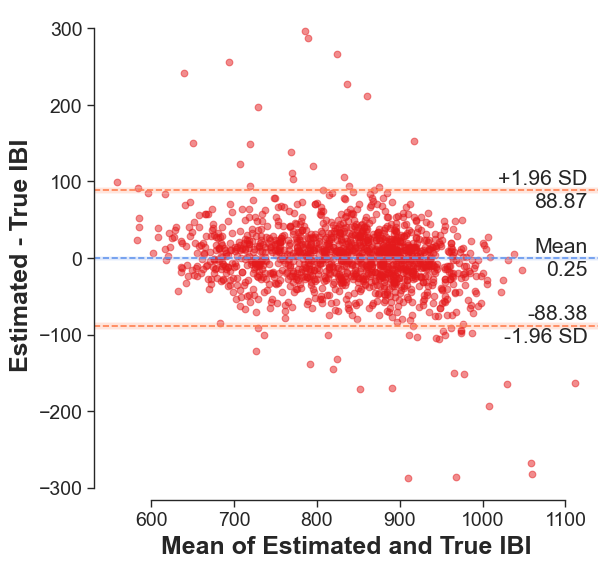}\\
         \subcaption{\textit{Sricharan et al.}}
         \label{BA6}
     \end{minipage}%
     \hfill
     \begin{minipage}[!htbp]{0.25\textwidth}
         \centering
         \includegraphics[width=\textwidth]{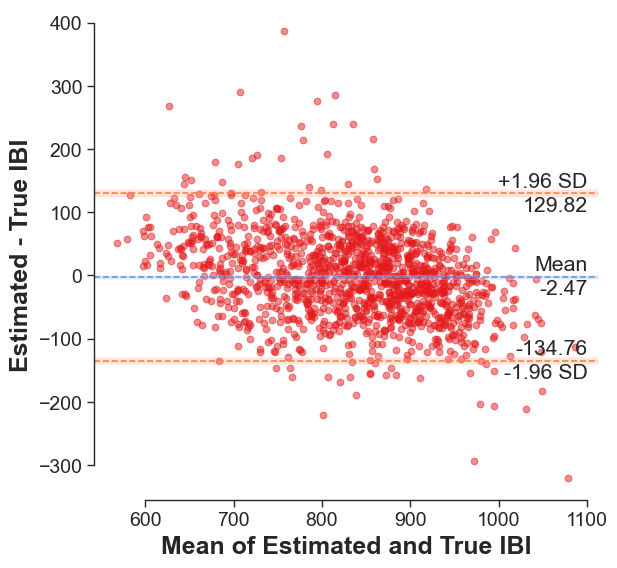}
         \subcaption{\textit{Aygun et al.}}
         \label{BA7}
     \end{minipage}%

     \begin{minipage}[!htbp]{0.25\textwidth}
         \centering
         \includegraphics[width=\textwidth]{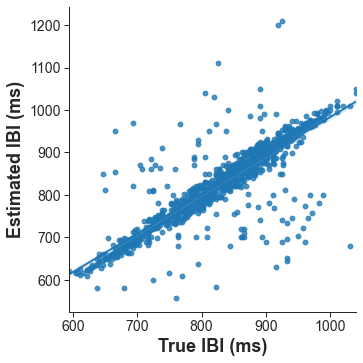}
         \subcaption{\textit{Brosnan et al.}\\\textit{r} = 0.8883}
         \label{C5}
     \end{minipage}%
     \hfill
     \begin{minipage}[!htbp]{0.25\textwidth}
         \centering
         \includegraphics[width=\textwidth]{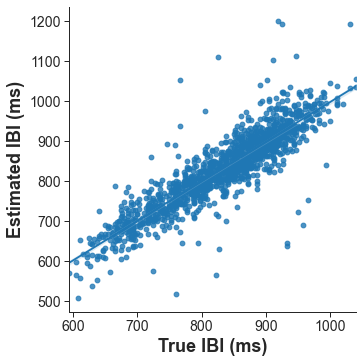}
         \subcaption{\textit{Sricharan et al.}\\\textit{r} = 0.8663}
         \label{C6}
     \end{minipage}%
     \hfill
     \begin{minipage}[!htbp]{0.25\textwidth}
         \centering
         \includegraphics[width=\textwidth]{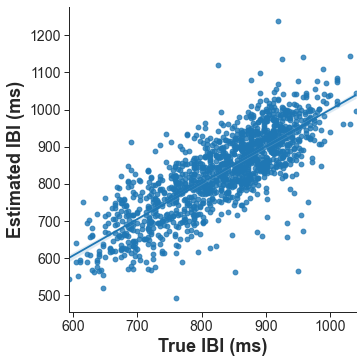}
         \subcaption{\textit{Aygun et al.}\\\textit{r} = 0.7897}
         \label{C7}
     \end{minipage}%

    \caption{Bland-Altman plot for IBI estimation on recording \# 230 using methods of \textit{Brosnan et al.}, \textit{Sricharan et al.}, and \textit{Aygun et al.} are depicted in Figs. \ref{BA5} to \ref{BA7} respectively. Corresponding correlation plots are also shown along with correlation coefficients in Figs. \ref{C5} to \ref{C7}.}
    \label{fig:six graphs}
\end{figure}

Prior studies in the field of ECG denoising and beat detection exercised \textit{SNR\textsubscript{imp}}, \textit{Sensitivity} or \textit{F\textsubscript{1} score} to evaluate their performance regarding detecting correct peaks. However, herein, we are interested in estimating IBI from noisy signals. Therefore, we selected our performance metrics based on that objective and leveraged \textit{Root Mean Squared Error (RMSE)} and \textit{\% error} of IBI for performance evaluation. We define those metrics as follows

\begin{align}
\operatorname { RMSE_{\textit{IBI}} } = \sqrt { \frac { \sum _ { i = 1 } ^ { n } ( P _ { i } - O _ { i } ) ^ { 2 } } { n } }
\end{align}

\begin{align}
\% error = \frac{1}{n}\sum _ { i = 1 } ^ { n }\frac { | O _ { i } - P _ { i } | } {O _ { i }  } \times 100\%
\end{align}
where, \textit{P\textsubscript{i}} refers to predicted IBI, \textit{O\textsubscript{i}} denotes observed IBI (groundtruth), and \textit{n} indicates number of IBIs in one segment of ECG.

The Bland-Altman plots in Figs. \ref{BA1} to \ref{BA4} show comparison of the IBIs estimated from noisy ECGs of SNR -30dB against their true IBIs. The ECGs considered in these plots are recordings \# 230, 220, 115 and 213. The limits of agreements (LOAs) - which work as boundary for 95\% of data and a good measurement of accuracy - are located at [49.92, -50.85] ms, [73.74, -72.44] ms, [20.14, -20.29] ms, and [33.89, -33.52] ms respectively. Also, the Pearson correlation plots are illustrated in \crefrange{C1}{C4} for the same set of recordings. The high correlation coefficients (\textit{r}) - 0.96, 0.93, 0.99, and 0.9 - emphasize that the estimated IBIs are very coherent to the true IBIs.

In comparison, similar plots are also shown in \crefrange{BA5}{BA7} and \crefrange{C5}{C7} highlighting the performance of \textit{Brosnan et al.}, \textit{Sricharan et al.}, and \textit{Aygun et al.} respectively on recording \# 230. The LOAs are at [90.25, -94.40] ms, [88.87, -88.38] ms, and [129.829, -134.76] ms respectively. So, our method is ensuring [40.33, -43.55] ms, [38.95, -37.53] ms, and [79.91, -83.91] ms of improvements respectively. As stated in corresponding papers, these methods also perform well under low noise situations. Correlation coefficients (\textit{r}) for these methods on this specific recording are 0.88, 0.87, and 0.79 respectively whereas our approach obtains 0.96 correlation for the same setting. For other recordings of same dataset, none of these 3 methods achieve correlation above 0.9 at an SNR value of -30dB. Performance of all methods (including ours) decline for recordings of IEEE SP Cup 2015 and European ST-T Database and the rationale has been clearly explained in Section \ref{disc}.

\begin{figure*}[!htbp]
\centering
\includegraphics[width = 0.80\textwidth]{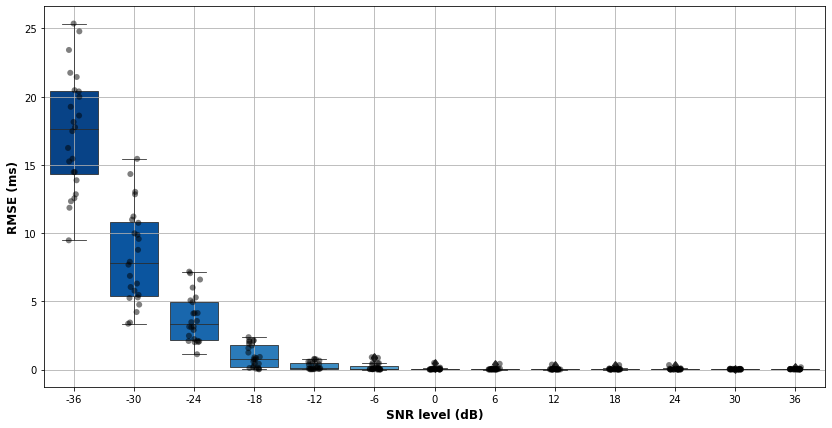}
\caption{Box and whisker plot for different SNR levels of test data of MIT-BIH Arrhythmia Database. On the x-axis we have SNR values starting from 36dB up to -36dB and on y-axis RMSE (ms) values are listed. The RMSEs are calculated by comparing the estimated IBIs with the true ones. Circular dots with low opacity are different RMSE values. The horizontal edges of colored boxes mark the interquartile range (IQR) (Q\textsubscript{1}-Q\textsubscript{3}) and contain 50\% data whereas the black horizontal lines inside each box mark the median value. The vertical whiskers hold other 50\% of the data and black horizontal lines mark the minimum and maximum. $\smblklozenge$ symbols are the outliers.} 
\label{fig:box}
\end{figure*}

The accuracy of the proposed methodology can also be inferred from the box and whisker plot depicted in Figure~\ref{fig:box}. The IBI values have been estimated for all 13 noisy variants of the test ECGs in MIT-BIH Arrhythmia Database listed in Table \ref{tab1} and compared with that of the clean versions. Notice that, we hardly get any error for SNR values up to -12dB and the estimated IBI values are consistent and nearly as good as the true IBIs. However, starting from -18dB, we encounter some amount of error and above -30dB the noise becomes very intense as well as the RMSE. The median value at -30dB is 7.792 ms.

\begin{figure}[!htbp]
\centering
\includegraphics[scale = 0.2]{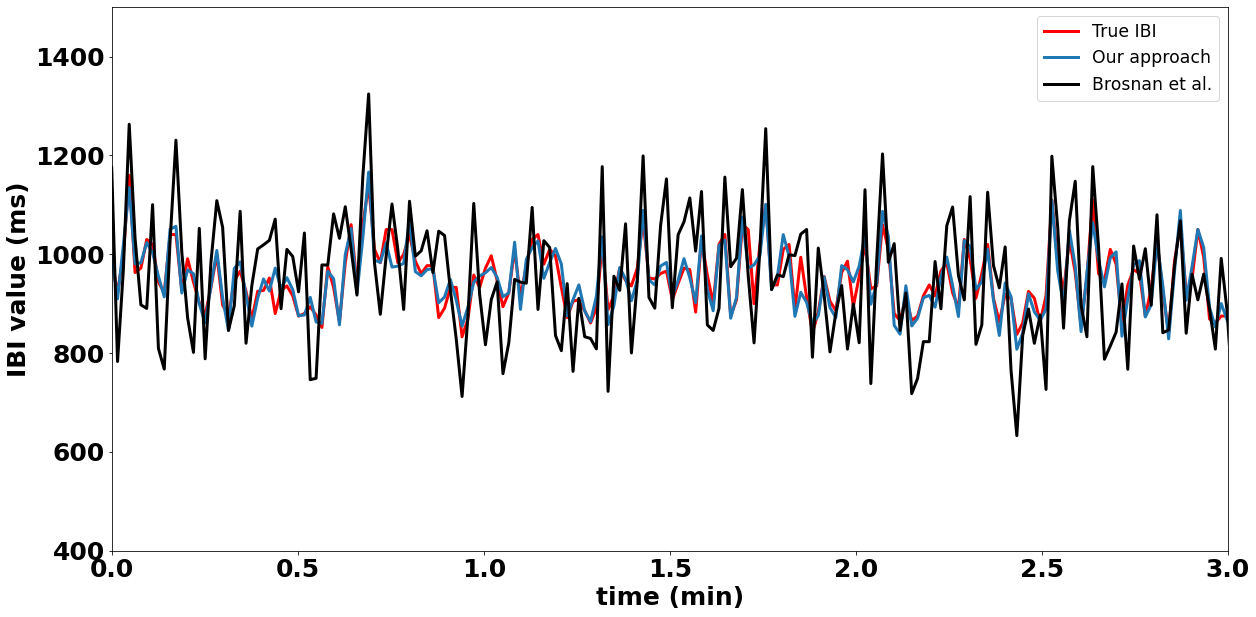}
\caption{Comparison of IBI plot with true IBIs and estimated IBIs obtained by \textit{Brosnan et al.}. For higher resolution, only first 3 minutes have been plotted.} 
\label{fig:IBI_comp}
\end{figure}

\begin{table*}[t]

\begin{subtable}[h]{\textwidth}
    \centering
\resizebox{0.8\textwidth}{!}{%
\begin{tabular}{p{0.18\linewidth}p{0.1\linewidth}p{0.1\linewidth}p{0.1\linewidth}p{0.1\linewidth}p{0.1\linewidth}p{0.1\linewidth}}
\hline
 & 0dB & -6dB & -12dB & -18dB & -24dB & -30dB \\
\hline
\multirow{2}{10em}{Brosnan et al.\cite{Yuen28}} & 1.69 ms & 6.86 ms & 15.41 ms & 21.22 ms & 28.73 ms & 31.44 ms \\
& 2.38\% & 6.97\% & 9.06\% & 12.74\% & 14.34\% & 17.24\% \\ \hline
\multirow{2}{10em}{Sricharan et al.\cite{Vijayarangan26}} & 2.64 ms & 8.61 ms & 12.02 ms & 20.05 ms & 49.78 ms & 54.33 ms \\
& 1.54\% & 3.79\% & 9.87\% & 18.71\% & 20.97\% & 22.41\% \\ \hline
\multirow{2}{10em}{Aygun et al.\cite{Aygun18}} & 5.08 ms & 8.41 ms & 17.24 ms & 32.47 ms & 42.32 ms & 55.72 ms \\ & 3.54\% & 5.88\% & 8.77\% & 14.27\% & 17.31\% & 21.83\% \\ \hline
\multirow{2}{10em}{Proposed Method} & 0.08 ms & 0.17 ms & 0.68 ms & 1.09 ms & 4.72 ms & 8.39 ms \\
& 0.01\% & 0.08\% & 0.28\%  & 0.95\% & 4.23\% & 6.34\% \\ \hline
    
    \end{tabular}}
    \caption{MIT-BIH Arrhythmia Database}
    \label{tab3}
\end{subtable}

\hfill

\begin{subtable}[h]{\textwidth}
    \centering
\resizebox{0.8\textwidth}{!}{%
\begin{tabular}{p{0.18\linewidth}p{0.1\linewidth}p{0.1\linewidth}p{0.1\linewidth}p{0.1\linewidth}p{0.1\linewidth}p{0.1\linewidth}}
 &  &  &  &  &  &  \\
\hline
 & 0dB & -6dB & -12dB & -18dB & -24dB & -30dB \\
\hline
\multirow{2}{10em}{Brosnan et al.\cite{Yuen28}} & 1.02 ms & 4.27 ms & 12.57 ms & 23.29 ms & 25.13 ms & 30.53 ms \\
& 1.87\% & 1.61\% & 2.59\% & 9.61\% & 15.02\% & 18.72\% \\ \hline
\multirow{2}{10em}{Sricharan et al.\cite{Vijayarangan26}} & 2.21 ms & 7.07 ms & 13.74 ms & 19.69 ms & 27.92 ms & 37.81 ms \\
& 1.21\% & 3.17\% & 11.51\% & 17.94\% & 21.38\% & 27.33\% \\ \hline
\multirow{2}{10em}{Aygun et al.\cite{Aygun18}} & 5.32 ms & 7.95 ms & 15.33 ms & 33.41 ms & 37.97 ms & 47.86 ms \\ & 3.21\% & 5.12\% & 7.91\% & 12.02\% & 18.32\% & 19.27\% \\ \hline
\multirow{2}{10em}{Proposed Method} & 2.27 ms & 3.85 ms & 4.09 ms & 5.76 ms & 7.98 ms & 10.52 ms \\
& 0.85\% & 1.32\% & 1.77\% & 2.31\% & 3.87\% & 7.71\% \\ \hline

    \end{tabular}}
    \caption{European ST-T Database}
    \label{tab4}
\end{subtable}

\hfill
\linebreak[1]
\begin{subtable}[h]{\textwidth}
    \centering
\resizebox{0.8\textwidth}{!}{%
\begin{tabular}{p{0.18\linewidth}p{0.1\linewidth}p{0.1\linewidth}p{0.1\linewidth}p{0.1\linewidth}p{0.1\linewidth}p{0.1\linewidth}}
 &  &  &  &  &  &  \\
\hline
 & 0dB & -6dB & -12dB & -18dB & -24dB & -30dB \\
\hline
\multirow{2}{10em}{Brosnan et al.\cite{Yuen28}} & 2.35 ms & 8.41 ms & 15.14 ms & 18.73 ms & 21.61 ms & 32.64 ms \\
& 2.17\% & 4.47\% & 7.37\% & 9.22\% & 13.36\% & 16.04\% \\ \hline
\multirow{2}{10em}{Sricharan et al.\cite{Vijayarangan26}} & 4.97 ms & 8.82 ms & 15.26 ms & 25.41 ms & 32.63 ms & 38.59 ms \\
& 1.73\% & 3.59\% & 9.85\% & 15.81\% & 17.23\% & 21.75\% \\ \hline
\multirow{2}{10em}{Aygun et al.\cite{Aygun18}} & 6.21 ms & 9.91 ms & 17.13 ms & 31.04 ms & 35.31 ms & 41.23 ms \\ & 3.64\% & 5.89\% & 7.42\%  & 11.90\% & 18.47\% & 18.98\% \\ \hline
\multirow{2}{10em}{Proposed Method} & 5.14 ms & 7.11 ms & 9.27 ms & 11.91 ms & 17.84 ms & 26.33 ms \\
& 3.63\% & 4.22\% & 6.96\% & 8.05\% & 8.79\% & 11.23\% \\ \hline

    \end{tabular}}
    \caption{IEEE Signal Processing Cup 2015}
    \label{tab5}
\end{subtable}

\caption{Comparison of different methods on 3 dataset. For any method, the upper value refers to avg. RMSE and the lower value refers to avg. \% error. Performance have been measured on signals listed in Table~\ref{tab1} only.}
\label{tab_comp}
\end{table*}

Plots of true IBIs, IBIs estimated with our approach and that of \textit{Brosnan et al.} (both at -30dB) are pictured in Figure~\ref{fig:IBI_comp}. Some other methods have not been included with a view to keeping the plots clean to understand. IBIs estimated with our approach are not that much different from the true one following the trend in a cogent way.

Lastly, we preform comparative analysis of 3 state-of-ther-art and our proposed method on MIT-BIH Arrhythmia Database v1.0.0, European ST-T Database, and IEEE Signal Processing Cup 2015 in ~\crefrange{tab3}{tab5} (the details of these methods are described in Section 2). Once more, the performance metrics are average RMSE and average \% error and noise has been added to all the signals from each dataset. In these tables we have considered SNR levels starting from 0dB to -30dB since performance of all methods on higher quality signals are almost flawless. The weighted average of RMSE is 13.514 ms and \% error is 7.97\% for the proposed method whereas they are 31.467 ms (17.953 ms higher than us) and 17.384\% (9.414\% higher) for \textit{Brosnan et al.}, 45.439 ms (31.925 ms higher) and 23.72\% (15.75\% higher) for \textit{Sricharan et al.}, 49.739 ms (36.225 ms higher) and 20.349\% (12.379\% higher) for \textit{Aygun et al.}. 

\section{Discussion}
\label{disc}

While analyzing the proposed method's performance, we have come across some of the observations, facts, and limitations which are worth mentioning-

The proposed algorithm can make the peaks more visible, detect them out of intense noisy signals and estimate the IBI values from it. However, it cannot redeem or rescue the typical shape of an ECG i.e. the PQRST wave. It can only recover the R-peaks.
    
Also, the IEEE SP Cup 2015 data is already encrypted with motion artifacts and we are adding additional noise to it. So, the unexpected performance on this dataset is quite reasonable.

It can be inferred from table \ref{tab_comp} that the performance of proposed algorithm on European ST-T Database and IEEE Signal Processing Cup 2015 is below than that of MIT-BIH Arrhythmia Database. Since the later is already sampled at 360Hz, we do not have to upsample it to add noise like the other two. So, upsampling of data degrades the performance on it.

\begin{figure}[h]
\centering
\includegraphics[scale = 0.225]{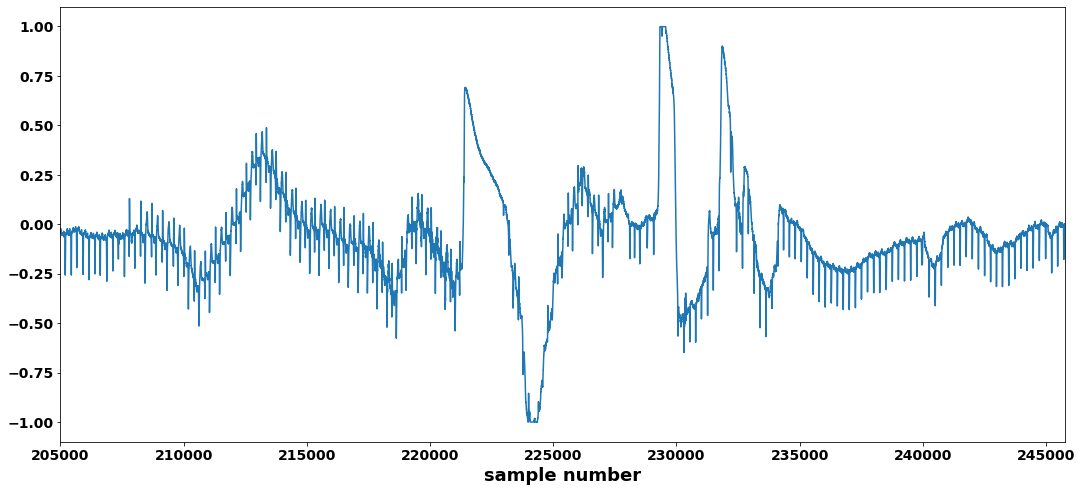}
\caption{A distorted ECG with missing peaks.} 
\label{dist}
\end{figure}
    
Some of the signals are not clean enough, missing peaks, distorted and have some innate noise. They are beyond the scope of fixing with any sort of pre processing. Such a distorted signal segment is presented in Figure~\ref{dist}. Obtaining IBIs from these types of signals is nearly impossible even in their pure form. So we had to neglect such signals from all database.
    
\begin{figure}[h]
\centering
\includegraphics[scale = 0.225]{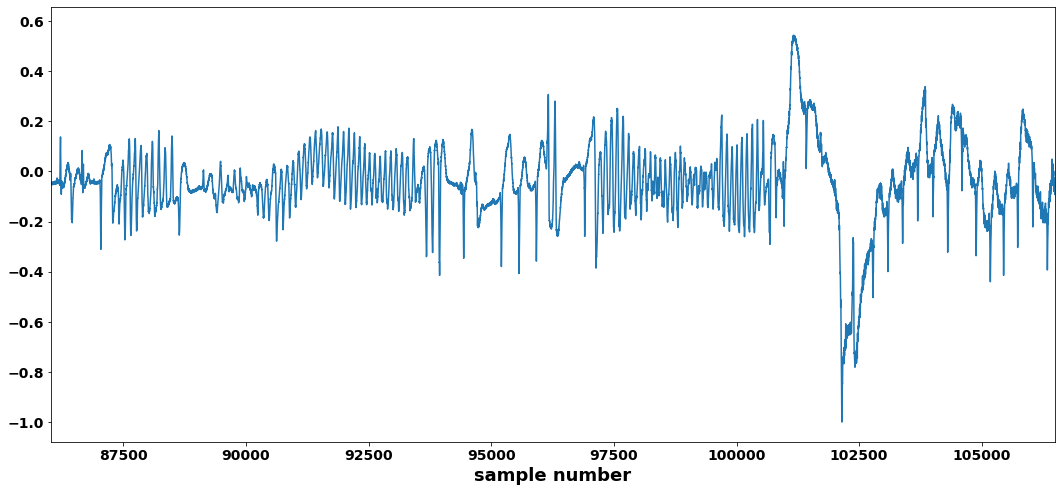}
\caption{An ECG with inconsistent peak size. There are too many such signals in each of the database. They have been omitted since they decrease the performance.} 
\label{inc_p}
\end{figure}

Some of the signals have inconsistent peak size, like the one in Figure~\ref{inc_p}. This becomes a notable hindrance. The SNR levels are set according to the power of the ECG and power of noise. Now, in an ECG with inconsistent peak size, the consistent peaks play dominant role to set the power of ECG signal and the noise power is set accordingly. In such a case, the noise peaks adjacent to the inconsistent peaks become much higher and the tiramisu model fails to differentiate between the actual ECG peak and the noise. So, it suppresses the actual peak and retains the noisy peak leading to higher error in IBI estimation.

Same goes for signals with SNR beyond -30dB. The noise signal gets so intense that it cannot be suppressed anymore. The detected peaks are not perfect and the IBI values estimated are far from accurate. So, we claim, our method is good for estimation up to -30dB.

\section{Conclusion}
\label{conc}
A stacked tiramisu model based novel method for extracting IBI from an ECG that is submerged in motion artifacts is presented and analyzed in this paper. This method can be easily used for this purpose and can compete with traditional concurrent techniques. The model simply suppresses the noise and later, the most prominent feature of ECG i.e. position of R-peaks, has been leveraged to complete the task. Also, it doesn't require any pre-processing or post-processing method which saves some of the complexities. Results show the efficiency of the proposed model and the estimated IBIs are highly correlated to that of the true IBIs even at -30dB SNR value. The weighted average RMSE for IBI estimation is 13.514 ms. Although the size of the model is massive, with some lighter versions of deep learning tools, it can also be made suitable for embedded system applications. On the other hand, this approach is easily understandable, handy, highly accurate, and bears the proof of practical feasibility. Also, with some slight changes, this method can also be used for other physiological signals. Undoubtedly, it is a deep learning based approach implemented for the betterment of mankind and we hope it will have far reaching effect in the field of cardiology since it can work as an early indicator of many cardiac diseases.

\bibliographystyle{unsrt}  
\bibliography{references}  

\begin{thebibliography}{10}

\bibitem{CDC_01}
Centre for Disease~Control and Prevention (CDC).
\newblock Heart disease facts.
\newblock Fact sheet, september 2020.

\bibitem{Aa01}
World Health~Organization (WHO).
\newblock Cardiovascular diseases (cvds).
\newblock Fact sheet, may 2017.

\bibitem{Ab02}
C.~McAloon, Luke~M Boylan, T.~Hamborg, N.~Stallard, F.~Osman, P.~Lim, and
  S.~Hayat.
\newblock The changing face of cardiovascular disease 2000-2012: An analysis of
  the world health organisation global health estimates data.
\newblock {\em International journal of cardiology}, 224:256--264, 2016.

\bibitem{Harju2018MonitoringOH}
J.~Harju, A.~Tarniceriu, Jakub Par{\'a}k, A.~Vehkaoja, A.~Yli-Hankala, and
  I.~Korhonen.
\newblock Monitoring of heart rate and inter-beat intervals with wrist
  plethysmography in patients with atrial fibrillation.
\newblock {\em Physiological measurement}, 39 6:065007, 2018.

\bibitem{Ahmed2010HeartRA}
Mobyen~Uddin Ahmed and S.~Begum.
\newblock Heart rate and inter-beat interval computation to diagnose stress
  using ecg sensor signal.
\newblock 2010.

\bibitem{Chessa2002RoleOH}
M.~Chessa, G.~Butera, G.~Lanza, E.~Bossone, A.~Delogu, G.~De Rosa, G.~Marietti,
  L.~Rosti, and M.~Carminati.
\newblock Role of heart rate variability in the early diagnosis of diabetic
  autonomic neuropathy in children.
\newblock {\em Herz}, 27:785--790, 2002.

\bibitem{Ac03}
C.~Varon, A.~Caicedo, D.~Testelmans, B.~Buyse, and S.~Huffel.
\newblock A novel algorithm for the automatic detection of sleep apnea from
  single-lead ecg.
\newblock {\em IEEE Transactions on Biomedical Engineering}, 62:2269--2278,
  2015.

\bibitem{Ad04}
M.~Bsoul, H.~Minn, and L.~Tamil.
\newblock Apnea medassist: Real-time sleep apnea monitor using single-lead ecg.
\newblock {\em IEEE Transactions on Information Technology in Biomedicine},
  15:416--427, 2011.

\bibitem{Ae05}
M.~Hasan, S.~Rahman, Asiful Arefeen, T.~Ahmed, Mohtasim Nakib, C.~Shahnaz, and
  Arik Subhana.
\newblock Portable real time ecg monitor and disease diagnostics.
\newblock {\em 2019 IEEE International Conference on Biomedical Engineering,
  Computer and Information Technology for Health (BECITHCON)}, pages 15--19,
  2019.

\bibitem{Af06}
Sheng Hu, H.~Wei, Y.~Chen, and J.~Tan.
\newblock A real-time cardiac arrhythmia classification system with wearable
  sensor networks.
\newblock {\em Sensors (Basel, Switzerland)}, 12:12844 -- 12869, 2012.

\bibitem{Qayyum2019ECGHC}
Alif Bin~Abdul Qayyum, Tanveerul Islam, and M.~Haque.
\newblock Ecg heartbeat classification: A comparative performance analysis
  between one and two dimensional convolutional neural network.
\newblock {\em 2019 IEEE International Conference on Biomedical Engineering,
  Computer and Information Technology for Health (BECITHCON)}, pages 93--96,
  2019.

\bibitem{Ag07}
Vu~Thi~Hong Nhan, N.~Park, Y.~K. Lee, Y.~Lee, J.~Lee, and K.~Ryu.
\newblock Online discovery of heart rate variability patterns in mobile
  healthcare services.
\newblock {\em J. Syst. Softw.}, 83:1930--1940, 2010.

\bibitem{Ah08}
Zhe Yang, Qihao Zhou, L.~Lei, K.~Zheng, and Wei Xiang.
\newblock An iot-cloud based wearable ecg monitoring system for smart
  healthcare.
\newblock {\em Journal of Medical Systems}, 40:1--11, 2016.

\bibitem{Islam2019SourceAC}
M.~F. Islam, Sheik Imran, Asiful Arefeen, M.~Hasan, and C.~Shahnaz.
\newblock Source and camera independent ophthalmic disease recognition from
  fundus image using neural network.
\newblock {\em 2019 IEEE International Conference on Signal Processing,
  Information, Communication \& Systems (SPICSCON)}, pages 59--63, 2019.

\bibitem{Chang2011GaussianNF}
K.~Chang and Shing-Hong Liu.
\newblock Gaussian noise filtering from ecg by wiener filter and ensemble
  empirical mode decomposition.
\newblock {\em Journal of Signal Processing Systems}, 64:249--264, 2011.

\bibitem{Kher15}
R.~Kher.
\newblock Signal processing techniques for removing noise from ecg signals.
\newblock 2019.

\bibitem{Tereshchenko2015FrequencyCA}
L.~Tereshchenko and M.~Josephson.
\newblock Frequency content and characteristics of ventricular conduction.
\newblock {\em Journal of electrocardiology}, 48 6:933--7, 2015.

\bibitem{Kirst2011UsingDF}
M.~Kirst, Bastian Glauner, and J.~Ottenbacher.
\newblock Using dwt for ecg motion artifact reduction with noise-correlating
  signals.
\newblock {\em 2011 Annual International Conference of the IEEE Engineering in
  Medicine and Biology Society}, pages 4804--4807, 2011.

\bibitem{Jgou29}
S.~J{\'e}gou, M.~Drozdzal, David V{\'a}zquez, A.~Romero, and Yoshua Bengio.
\newblock The one hundred layers tiramisu: Fully convolutional densenets for
  semantic segmentation.
\newblock {\em 2017 IEEE Conference on Computer Vision and Pattern Recognition
  Workshops (CVPRW)}, pages 1175--1183, 2017.

\bibitem{Pandey09}
V.~Pandey and V.~K. Giri.
\newblock High frequency noise removal from ecg using moving average filters.
\newblock {\em 2016 International Conference on Emerging Trends in Electrical
  Electronics \& Sustainable Energy Systems (ICETEESES)}, pages 191--195, 2016.

\bibitem{Almal10}
Mohammed~Tali Almalchy, V.~Ciobanu, and N.~Popescu.
\newblock Noise removal from ecg signal based on filtering techniques.
\newblock {\em 2019 22nd International Conference on Control Systems and
  Computer Science (CSCS)}, pages 176--181, 2019.

\bibitem{Rakshit11}
Hrishi Rakshit and M.~A. Ullah.
\newblock A new efficient approach for designing fir low-pass filter and its
  application on ecg signal for removal of awgn noise.
\newblock 2016.

\bibitem{Karagiannis12}
A.~Karagiannis and P.~Constantinou.
\newblock On the empirical mode decomposition performance in white gaussian
  noise biomedical signals.
\newblock 2010.

\bibitem{Talbi13}
M.~Talbi, S.~Abid, and A.~Cherif.
\newblock Emd-based ecg denoising using source separation.
\newblock {\em Journal of Mechanics in Medicine and Biology}, 15:1550082, 2015.

\bibitem{Pan1985ARQ}
J.~Pan and W.~Tompkins.
\newblock A real-time qrs detection algorithm.
\newblock {\em IEEE Transactions on Biomedical Engineering}, BME-32:230--236,
  1985.

\bibitem{Sundararaj14}
Vinu Sundararaj.
\newblock An efficient threshold prediction scheme for wavelet based ecg signal
  noise reduction using variable step size firefly algorithm.
\newblock {\em International Journal of Intelligent Engineering and Systems},
  9:117--126, 2016.

\bibitem{GilgenAmmann16}
Rahel Gilgen-Ammann, Theresa Schweizer, and T.~Wyss.
\newblock Rr interval signal quality of a heart rate monitor and an ecg holter
  at rest and during exercise.
\newblock {\em European Journal of Applied Physiology}, 119:1525--1532, 2019.

\bibitem{Rankawat22}
Shalini~A. Rankawat and R.~Dubey.
\newblock Robust heart rate estimation from multimodal physiological signals
  using beat signal quality index based majority voting fusion method.
\newblock {\em Biomed. Signal Process. Control.}, 33:201--212, 2017.

\bibitem{Rezk19}
S.~Rezk, C.~Join, and S.~E. Asmi.
\newblock Inter-beat (r-r) intervals analysis using a new time delay estimation
  technique.
\newblock {\em 2012 Proceedings of the 20th European Signal Processing
  Conference (EUSIPCO)}, pages 929--933, 2012.

\bibitem{Aygun17}
Ayca Aygun and R.~Jafari.
\newblock Robust heart rate variability and interbeat interval detection
  algorithm in the presence of motion artifacts.
\newblock {\em 2019 IEEE EMBS International Conference on Biomedical \& Health
  Informatics (BHI)}, pages 1--5, 2019.

\bibitem{Aygun18}
Ayca Aygun, H.~Ghasemzadeh, and R.~Jafari.
\newblock Robust interbeat interval and heart rate variability estimation
  method from various morphological features using wearable sensors.
\newblock {\em IEEE Journal of Biomedical and Health Informatics},
  24:2238--2250, 2020.

\bibitem{Pulavskyi20}
A.~Pulavskyi, S.~Krivenko, and L.~S. Kryvenko.
\newblock Evaluation of the effectiveness of post-filtration smoothing using
  lossless compression for heart rate variability obtained from a very noisy
  ecg.
\newblock {\em 2020 9th Mediterranean Conference on Embedded Computing (MECO)},
  pages 1--5, 2020.

\bibitem{Ansari21}
S.~Ansari, Jonathan Gryak, and K.~Najarian.
\newblock Noise detection in electrocardiography signal for robust heart rate
  variability analysis: A deep learning approach.
\newblock {\em 2018 40th Annual International Conference of the IEEE
  Engineering in Medicine and Biology Society (EMBC)}, pages 5632--5635, 2018.

\bibitem{Laitala23}
Juho Laitala, Mingzhe Jiang, Elise Syrj{\"a}l{\"a}, Emad~Kasaeyan Naeini, Antti
  Airola, A.~Rahmani, N.~Dutt, and P.~Liljeberg.
\newblock Robust ecg r-peak detection using lstm.
\newblock {\em Proceedings of the 35th Annual ACM Symposium on Applied
  Computing}, 2020.

\bibitem{Reljin24}
N.~Reljin, J.~L{\'a}zaro, Md~Billal Hossain, Yeon~Sik Noh, C.~Cho, and K.~Chon.
\newblock Using the redundant convolutional encoder–decoder to denoise qrs
  complexes in ecg signals recorded with an armband wearable device.
\newblock {\em Sensors (Basel, Switzerland)}, 20, 2020.

\bibitem{Ronneberger25}
O.~Ronneberger, P.~Fischer, and T.~Brox.
\newblock U-net: Convolutional networks for biomedical image segmentation.
\newblock {\em ArXiv}, abs/1505.04597, 2015.

\bibitem{Vijayarangan26}
Sricharan Vijayarangan, Vignesh Ravichandran, Balamurali Murugesan,
  S.~Preejith, J.~Joseph, and M.~Sivaprakasam.
\newblock Rpnet: A deep learning approach for robust r peak detection in noisy
  ecg.
\newblock {\em 2020 42nd Annual International Conference of the IEEE
  Engineering in Medicine \& Biology Society (EMBC)}, pages 345--348, 2020.

\bibitem{Qiu27}
Lishen Qiu, Wenqiang Cai, Jie Yu, J.~Zhong, Yan Wang, Wanyue Li, Y.~Chen, and
  L.~Wang.
\newblock A two-stage ecg signal denoising method based on deep convolutional
  network.
\newblock {\em bioRxiv}, 2020.

\bibitem{Antczak2020AGA}
K.~Antczak.
\newblock A generative adversarial approach to ecg synthesis and denoising.
\newblock {\em ArXiv}, abs/2009.02700, 2020.

\bibitem{Yuen28}
B.~Yuen, X.~Dong, and Tao Lu.
\newblock Detecting noisy ecg qrs complexes using waveletcnn autoencoder and
  convlstm.
\newblock {\em IEEE Access}, 8:143802--143817, 2020.

\bibitem{Moody2001TheIO}
G.~Moody and R.~Mark.
\newblock The impact of the mit-bih arrhythmia database.
\newblock {\em IEEE Engineering in Medicine and Biology Magazine}, 20:45--50,
  2001.

\bibitem{Goldberger2000PhysioBankPA}
A.~Goldberger, L.~Amaral, L.~Glass, Jeffrey~M. Hausdorff, P.~Ivanov, R.~Mark,
  J.~Mietus, G.~Moody, C.~Peng, and H.~Stanley.
\newblock Physiobank, physiotoolkit, and physionet: components of a new
  research resource for complex physiologic signals.
\newblock {\em Circulation}, 101 23:E215--20, 2000.

\bibitem{Taddei1992TheES}
A.~Taddei, G.~Distante, M.~Emdin, P.~Pisani, G.~Moody, C.~Zeelenberg, and
  C.~Marchesi.
\newblock The european st-t database: standard for evaluating systems for the
  analysis of st-t changes in ambulatory electrocardiography.
\newblock {\em European heart journal}, 13 9:1164--72, 1992.

\bibitem{Zhang2015TROIKAAG}
Zhilin Zhang, Zhouyue Pi, and B.~Liu.
\newblock Troika: A general framework for heart rate monitoring using
  wrist-type photoplethysmographic signals during intensive physical exercise.
\newblock {\em IEEE Transactions on Biomedical Engineering}, 62:522--531, 2015.

\bibitem{Moody1992TheMN}
G.~Moody, We~Muldrow, and R.~Mark.
\newblock The mit-bih noise stress test database.
\newblock 1992.

\bibitem{peak}
Eric Jones, Travis Oliphant, Pearu Peterson, et~al.
\newblock {SciPy}: Open source scientific tools for {Python}, 2001--.

\bibitem{American_Health}
American~Heart Association.
\newblock All about heart rate (pulse), july 2015.

\bibitem{Athletes}
Jane Chertoff.
\newblock Why do athletes have a lower resting heart rate?
\newblock Fact sheet, april 2020.

\end{thebibliography}

\section{Online Resources}

The code for this paper can be found at: \textcolor{blue}{\url{https://github.com/Arefeen06088/IBI_Tiramisu}}

\end{document}